\documentclass[apl,
reprint,
superscriptaddress]{revtex4-1}

\usepackage{graphicx}
\usepackage{amsmath}
\usepackage[]{natbib}

\usepackage[colorlinks=true,urlcolor=black,citecolor=black, linkcolor=black]{hyperref}

\usepackage{textcomp}

\begin{document}

\title{Valence state of Sm in single crystalline EuO thin films}

\author{A.~Reisner}
\affiliation{Max Planck Institute for Chemical Physics of Solids, N\"othnitzer Str. 40, 01187 Dresden, Germany}

\author{D.~Kasinathan}
\affiliation{Max Planck Institute for Chemical Physics of Solids, N\"othnitzer Str. 40, 01187 Dresden, Germany}

\author{S.~Wirth}
\affiliation{Max Planck Institute for Chemical Physics of Solids, N\"othnitzer Str. 40, 01187 Dresden, Germany}

\author{L.~H.~Tjeng}
\affiliation{Max Planck Institute for Chemical Physics of Solids, N\"othnitzer Str. 40, 01187 Dresden, Germany}

\author{S.~G.~Altendorf}
\affiliation{Max Planck Institute for Chemical Physics of Solids, N\"othnitzer Str. 40, 01187 Dresden, Germany}


\begin{abstract}
Samarium has two stable valence states, 2+ and 3+, which coexist in many compounds forming spatially homogeneous intermediate valence states. We study the valence state of samarium when incorporated in a single crystalline EuO thin film which crystallizes in a $fcc$-structure similar to that of the intermediate valence SmO, but with a larger lattice constant. Due to the increased lattice spacing, a stabilization of the larger Sm$^{2+}$ ion is expected. Surprisingly, the samarium incorporated in Sm$_{\mathrm{x}}$Eu$_{\mathrm{1-x}}$O thin films shows a predominantly trivalent character, as determined by x-ray photoelectron spectroscopy and magnetometry measurements. We infer that the O$^{2-}$ ions in the EuO lattice have enough room to move locally, so as to reduce the Sm-O distance and stabilize the Sm$^{3+}$ valence. (Dated: April 06, 2017)
\end{abstract}

%\begin{document}

\maketitle
\section{Introduction}
Interest in the study of samarium containing compounds has increased recently, owing to the theoretical prediction of topologically nontrivial surface states in SmB$_{6}$ \cite{dzero10, takimoto11}, the high-pressure golden phase of SmS \cite{li14}, and ambient SmO \cite{kasinathan15}. The speciality of samarium arises from the perpetual competition between its two stable valence states: nonmagnetic Sm$^{2+}$ and a magnetic Sm$^{3+}$, which is stabilized in many solids by the formation of spatially homogenous non-integral intermediate valence Sm$^{2.{\mathrm{x}}+}$. Experimental examples of samarium containing intermediate valence compounds include the extensively studied SmB$_{6}$ \cite{vainshtein1965, menth1969}; high-pressure golden-phases of SmS, SmSe and SmTe \cite{sidorov1989}; ambient SmO \cite{yacoubi1979}; SmFe$_{2}$Al$_{10}$ \cite{peratheepan2015}; Sm$_{2.75}$C$_{60}$ \cite{arvanitidis2003}; and elemental samarium metal itself \cite{wertheim1977}.
The intermediate valence state of samarium can be understood as an intra-atomic promotion of 4$f$ electrons to 5$d$ bands, going from Sm$^{2+}$ (4$f^{6}$5$d^{0}$) to Sm$^{3+}$ (4$f^{5}$5$d^{1}$). Combined with the effects of strong electronic correlations and strong spin-orbit coupling, various exotic ground states are realizable in samarium based materials.

\begin{figure*}
	\centering
		\includegraphics[width=1\textwidth]{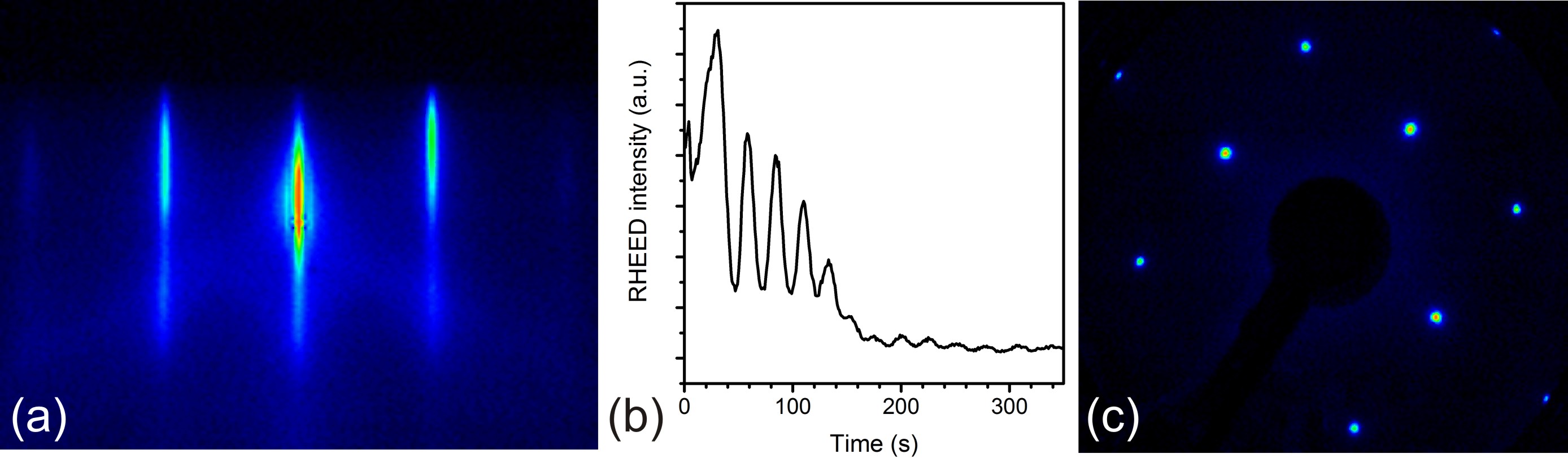}
				\caption{(a) RHEED image of the EuO thin film with 9\% Sm. (b) corresponding RHEED intensity oscillations of the specularly reflected electron beam. (c) LEED pattern of the EuO film with 9\% Sm recorded at 61 eV.}
	\label{fig:structure}
\end{figure*}

In this work we select the SmO and EuO system and address the question of the valence of Sm in single crystalline EuO films. Our interest is motivated by the proposal \cite{kasinathan15} that SmO is a candidate material for having topologically protected surface states and that a thin film interface of SmO with the ferromagnetic semiconductor EuO may make an excellent contender towards realizing the quantum anomalous Hall effect in strongly correlated electron systems. While the preparation of high quality single crystalline EuO thin films is well established \cite{sutarto09a}, no report exists to our knowledge about SmO films. In fact, the synthesis of SmO as bulk material requires high pressures \cite{leger80, krill80}, suggesting that the preparation of SmO in thin film form may not be possible using standard thin film deposition techniques. As a start, we are therefore interested to make a system containing SmO$_6$ octahedral units by utilizing EuO as the stabilizing host material, i.e. to prepare Sm$_{\mathrm{x}}$Eu$_{\mathrm{1-x}}$O, and study the local properties of the Sm.

Both SmO and EuO crystallize in a face centered cubic ($fcc$) lattice with comparable lattice constants of
4.9414 \AA \ \cite{krill80} and 5.1435 \AA \ \cite{McMasters74}, respectively. Based on empirical estimates of the lattice constants for Sm$^{3+}$O (4.917 \AA) and Sm$^{2+}$O (5.15 \AA), \cite{leger80} the samarium valency in bulk SmO was determined to be 2.9+, not inconsistent with x-ray absorption spectroscopy measurements \cite{krill80}. EuO on the other hand consists of the integer valent and magnetic Eu$^{2+}$, and exhibits many spectacular properties including a nearly 100\% spin polarized conduction band \cite{steeneken02, schmehl07}, a metal-to-insulator transition (MIT) \cite{torrance72} and  a colossal magneto-resistance  \cite{shapira73} in the slightly Eu-rich compound, as well as large magneto-optical effects \cite{ahn70b, wang86}.
We utilize the well established Eu distillation technique \cite{sutarto09a} to grow high-quality single crystalline Sm$_{\mathrm{x}}$Eu$_{\mathrm{1-x}}$O thin films by molecular beam epitaxy (MBE) under ultra high vacuum conditions. We expect that the larger EuO lattice will allow the Sm to be stabilized in the 2+ valence state or at least in an intermediate valence state closer to 2+ compared to bulk SmO.\\

\section{Experimental Methods}

Thin films of Sm$_{\mathrm{x}}$Eu$_{\mathrm{1-x}}$O were grown in a MBE system with a base pressure of 1$\times$10$^{-10}$ mbar. Prior to growth, epi-polished yttria stabilized zirconia (YSZ) substrates with a (001) surface (provided by CrysTec GmbH) were annealed for two hours at 600 \textcelsius $ $ in an oxygen atmosphere of 5$\times$10$^{-7}$ mbar. High purity Eu metal (Ames Laboratory) and Sm metal (Hunan HMC Metals C., Ltd.) were co-evaporated from effusion cells (LUXEL RADAK I). The Sm was additionally purified by degassing at about 1000 \textcelsius \ in a Mo ampulla before use.

Eu was evaporated using a flux rate of 8.0 $\pm$ 0.2 \AA/min at temperatures of about 570 \textcelsius. Sm was evaporated at flux rates of 0.02 - 1.5 \AA/min at temperatures between 450 - 520 \textcelsius.
Molecular oxygen was supplied via a high precision leak valve and set to a partial pressure of about 3.0$\times$10$^{-8}$ mbar during film growth. The YSZ substrates were heated to 400 \textcelsius $ $ during deposition. These growth parameters were calibrated to fulfill the Eu distillation condition \cite{sutarto09a}.
The growth was monitored by reflection high energy electron diffraction (RHEED) using a STAIB Instruments EK-35-R system. The diffraction patterns were recorded at an electron energy of 15 kV. After 60 minutes, the growth was terminated by first closing the oxygen supply and the Sm shutter at the same time. To avoid a formation of Eu$^{3+}$ at the film surface, after further 30 seconds the Eu shutter was closed and the substrate cooled down to room temperature.
Two reference films were additionally prepared: i) intermediate valence metallic Sm was deposited at room temperature onto an annealed YSZ substrate, ii) a trivalent Sm$_2$O$_3$ film was grown at room temperature onto a Sm film at a Sm evaporation rate of 7.1 \AA /min in an oxygen partial pressure of 1$\times$10$^{-7}$ mbar. \\

\begin{table}
\begin{center}
\begin{tabular}{|c|c||c|c|}
	\hline	
\multicolumn{2}{|l||}{Sm concentration (\%)} &\multicolumn{2}{l|}{Thickness (nm)}	 \\						 
from XPS	&	from flux rate			&	EuO	&	Cr	\\
													\hline	
-		&	2.4					&	 39 	&	13	\\
5.2		&	5.2					&	38		&	11	\\
9.1		&	9.8					&	38		&	12	\\
12.0		&	11.9					&	43		&	12	\\
17.5		&	13.0					&	43		&	9	\\
19.9		&	20.0					&	43		&	10\\	
	\hline	
\end{tabular}
\end{center}
\caption{Overview of Sm concentrations as derived from x-ray photoelectron spectra and Eu/Sm flux rates (see text), and layer thicknesses as deduced from x-ray reflectivity measurements. Cr was used for capping.}
\label{tab:values}
\end{table}

\begin{figure*}
	\centering
		\includegraphics[width=1\textwidth]{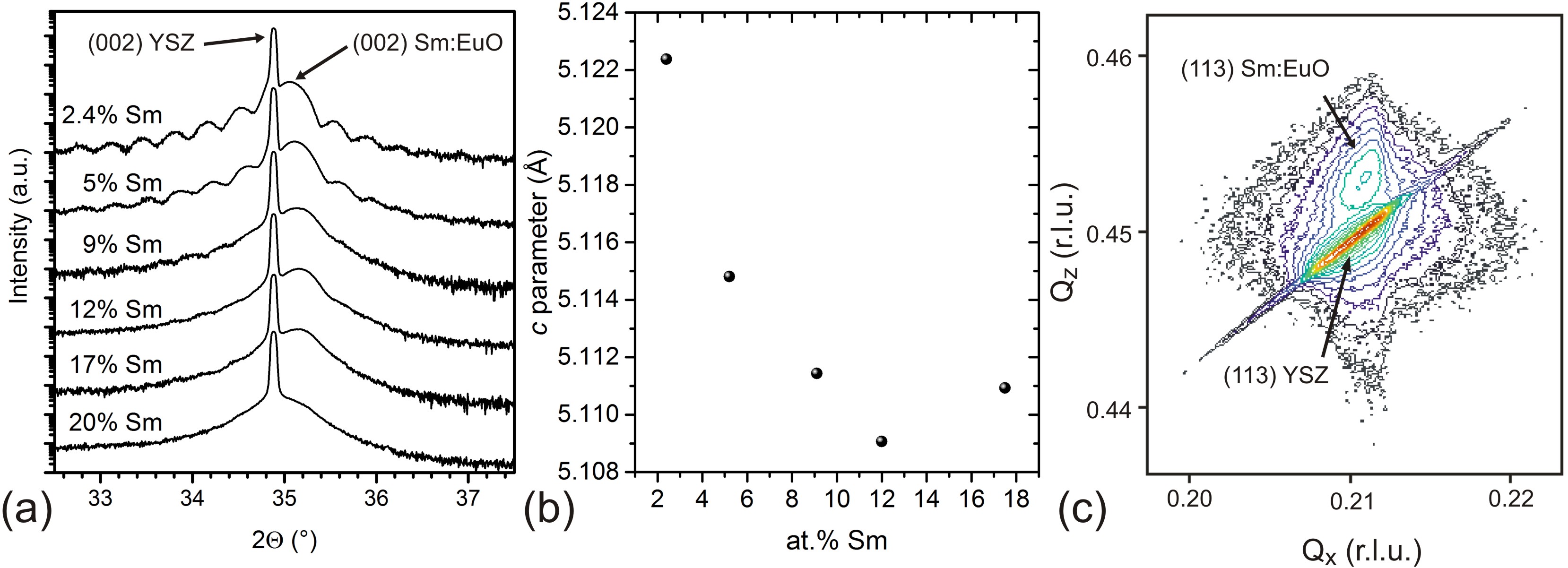}
	\caption{(a) $\theta$-2$\theta$ XRD scans around the (002) reflection of the Sm$_{\mathrm{x}}$Eu$_{\mathrm{1-x}}$O films. (b) The corresponding out-of-plane ($c$) lattice parameters calculated from the 2$\theta$ value as a function of Sm concentration. (c) XRD reciprocal space map around the (113) reflection of the EuO film with 17\%  Sm.}
	\label{fig:XRD}
\end{figure*}

The films were analyzed \textit{in situ} by x-ray photoelectron spectroscopy (XPS), utilizing a monochromatized Al $K_{\alpha}$ x-ray source (h$\nu$ = 1486.6 eV) and a Scienta R3000 electron energy analyzer. Spectra were collected at room temperature and in normal emission geometry. Subsequently, low energy electron diffraction (LEED) measurements were performed using a Thermo VG Scienta T191 system.
For further \textit{ex situ} characterization, the films were covered with a capping layer of about 100 \AA \ thick Cr to prevent the sensitive EuO from degradation in air.
X-ray diffraction (XRD), x-ray reflectivity (XRR) measurements and reciprocal space maps (RSM) were recorded in an X'Pert PRO four circle diffractometer from PANalytical equipped with a monochromatized Cu $K_{\alpha 1}$ radiation source and a PIXcel-3D detector.
The magnetic properties were investigated using a Quantum Design MPMS XL5 magnetometer. The magnetization curves were measured by applying an in-plane magnetic field of 50 Oe during the temperature sweeps.
Details of the samples - Sm concentrations and layer thicknesses - are summarized in table \ref{tab:values}.\\

\section{Results}
The crystalline quality of the Sm$_{\mathrm{x}}$Eu$_{\mathrm{1-x}}$O thin films is confirmed by the sharp streaks of the RHEED patterns for the entire substitution range reported. An exemplary electron diffraction image of the EuO sample with 9\% Sm is shown in fig. \ref{fig:structure} (a). The intensity oscillations of the specularly reflected electron beam reveal a smooth layer-by-layer growth (fig. \ref{fig:structure} (b)). One can distinguish two different growth mechanisms which are characteristic for the deposition of EuO on YSZ, cf. Ref. \cite{sutarto09a}. In the initial growth, oxygen is supplied mainly from the YSZ substrate yielding the first pronounced intensity oscillations typical for the growth of EuO on YSZ. After the completion of five monolayers, the oxygen is supplied only through the partial oxygen pressure in the growth chamber. The LEED pattern in fig.  \ref{fig:structure} (c) reflects the fourfold symmetry of the (001) facet of the \textit{fcc} rock salt structure and indicates a well ordered single crystalline surface.\\

\begin{figure}
	\centering
		\includegraphics[width=1\columnwidth]{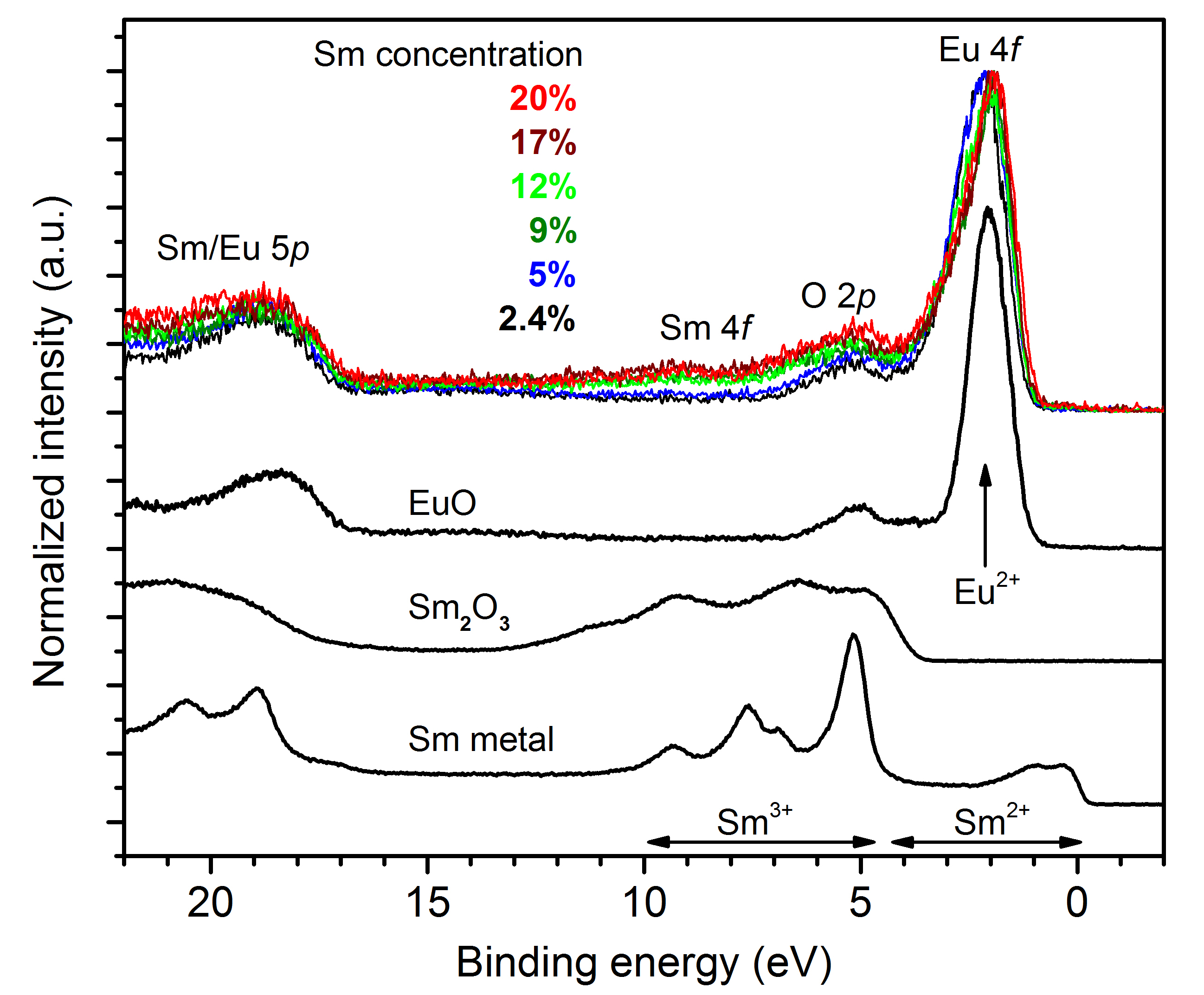}
	\caption{XPS spectra of the valence band region of the Sm$_{\mathrm{x}}$Eu$_{\mathrm{1-x}}$O thin films for various Sm concentrations up to 20\%. For comparison, the spectra of a pure EuO film, a Sm$_2$O$_3$  thin film and a pure Sm metal film are shown additionally. The contributions of Sm$^{3+}$ and Sm$^{2+}$ to the intermediate valence Sm metal spectrum are indicated. \cite{strisland96}}
	\label{fig:VB}
\end{figure}

The structure of the Sm$_{\mathrm{x}}$Eu$_{\mathrm{1-x}}$O thin films was analyzed in more detail by \textit{ex situ} XRD measurements (fig. \ref{fig:XRD}). The $\theta$-2$\theta$ scans show the sharp (002) peaks of the YSZ substrate and, at slightly higher diffraction angles, the broader (002) peaks of the Sm$_{\mathrm{x}}$Eu$_{\mathrm{1-x}}$O films with interference fringes, a signature of the finite layer thickness. With increasing Sm substitution concentration, the film peaks shift to larger 2$\theta$ diffraction angles, i.e., the  out-of-plane ($c$) lattice constant decreases. The deduced lattice parameters are summarized in fig. \ref{fig:XRD} (b).  The thickness fringes vanish for high Sm substitution concentrations indicating an increasing disorder and roughness.
The RSM analysis reveals that the in-plane lattice parameter of Sm$_{\mathrm{x}}$Eu$_{\mathrm{1-x}}$O adapts to the YSZ substrate such that the grown films are coherently strained up to a Sm substitution level of 17\%, see fig. \ref{fig:XRD} (c). This is not inconsistent with the reported solubility of 14 at.\% Sm in EuO bulk samples \cite{samokhvalov76}. In fact, our observation that the out-of-plane lattice constant of the 17\% sample does not follow the decreasing trend may indeed indicate that we have exceeded the solubility here.

\begin{figure}
	\centering
		\includegraphics[width=1\columnwidth]{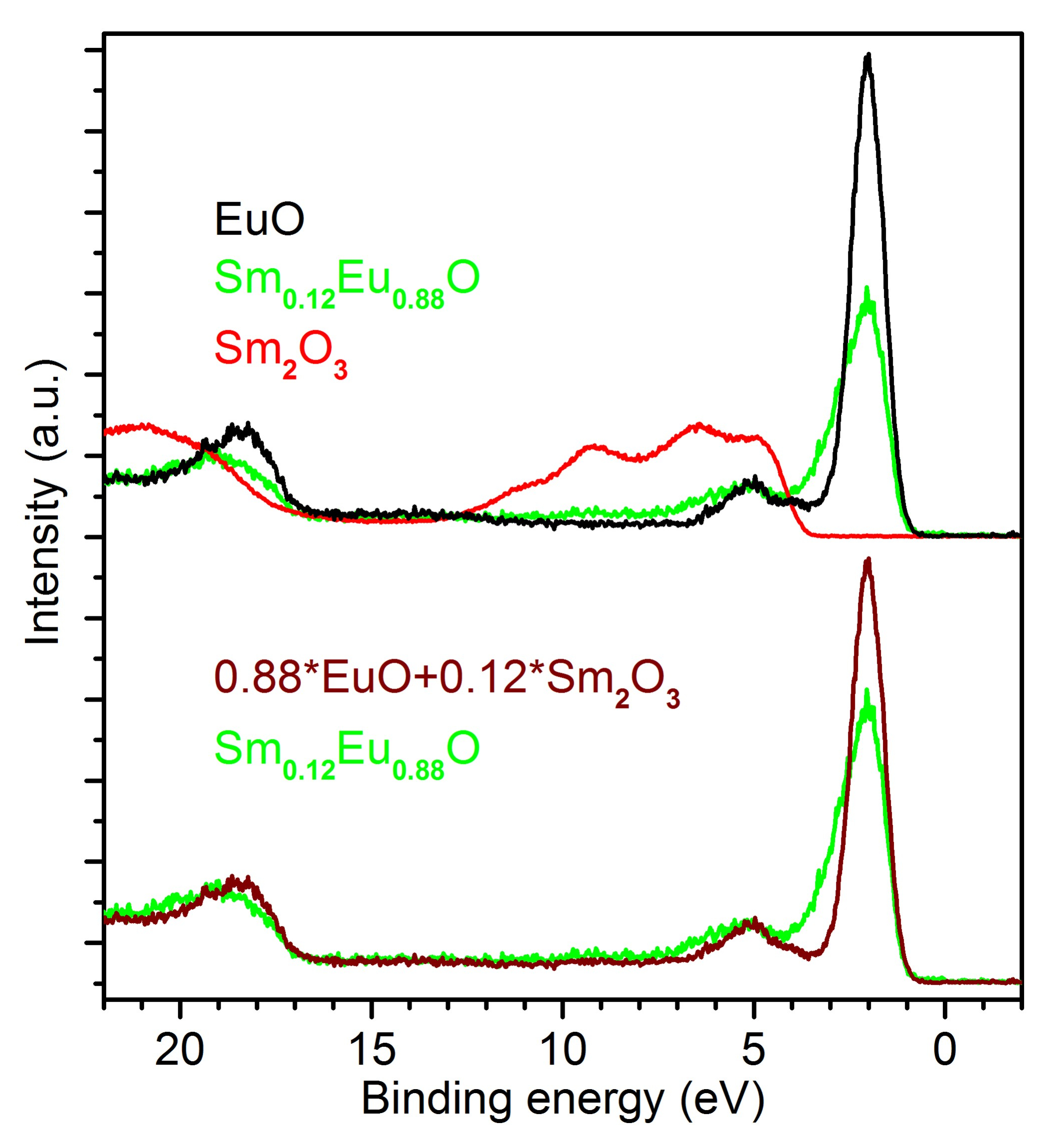}
	\caption{\textit{top}: XPS spectra of the valence band region of the pure EuO, the EuO with 12\% Sm, and a Sm$_2$O$_3$  thin film. \textit{bottom}: A comparison of a reference spectrum composed of 12\% of the Sm$_2$O$_3$ spectrum and 88\% of the EuO spectrum, and the measured spectrum of the EuO thin film with 12\% Sm.}
	\label{fig:VBcompo}
\end{figure}

\begin{figure}
	\centering
		\includegraphics[width=1.00\columnwidth]{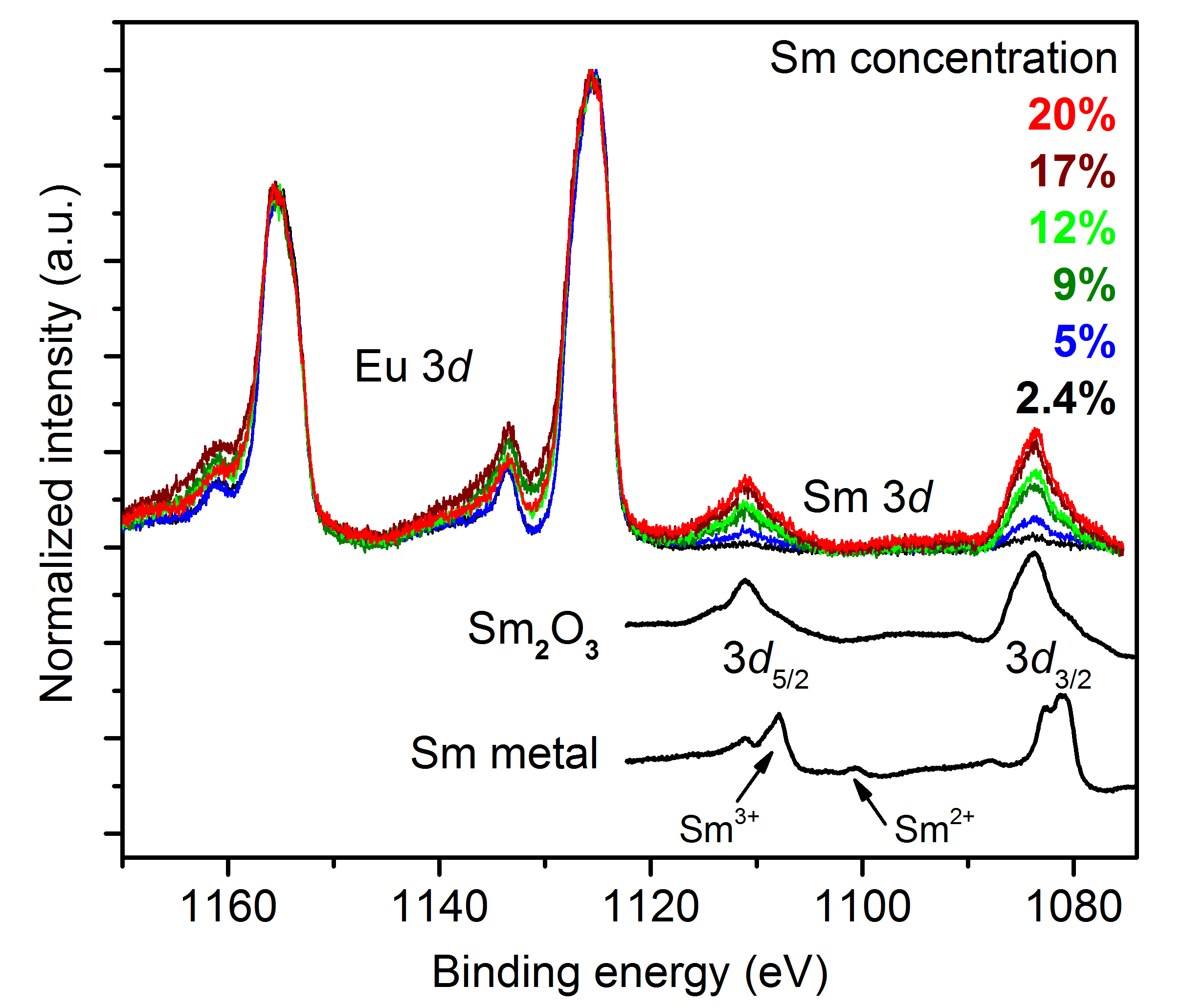}
	\caption{XPS spectra of the Eu 3\textit{d} and Sm 3\textit{d} core level regions of the Sm$_{\mathrm{x}}$Eu$_{\mathrm{1-x}}$O thin films. For comparison the spectra of a Sm$_2$O$_3$  thin film and a pure Sm metal film are shown additionally. The energy position of the Sm$^{3+}$ and Sm$^{2+}$ contributions to the 3\textit{d}$_{5/2}$ level of Sm metal are indicated. \cite{wertheim78}}
	\label{fig:3d}
\end{figure}

To check the chemical composition and the valence states of the constituents, \textit{in situ} XPS measurements were performed. The valence of the Eu atoms in Sm$_{\mathrm{x}}$Eu$_{\mathrm{1-x}}$O films can be directly deduced from the valence band scans shown in fig. \ref{fig:VB}. The 4$f^{7}\to4f^{6}$ transition at about 2 eV binding energy corresponds to the Eu$^{2+}$ valence state. The absence of the Eu$^{3+}$ peak at 7 eV confirms the purely divalent character of the Eu.
With increasing Sm content, the Sm 4$f$ states can be recognized in the range of 4-10 eV binding energy. A comparison with the trivalent Sm$_2$O$_3$ reference indicates that the Sm valence in the films is predominantly Sm$^{3+}$.
In fig. \ref{fig:VBcompo}, we have simulated the spectrum of Sm$_{\mathrm{0.12}}$Eu$_{\mathrm{0.88}}$O by composing a spectrum consisting of 12\% trivalent Sm$_2$O$_3$ and 88\% divalent EuO. The spectral weight around 5-10 eV binding energy is reproduced quite well, implying the trivalent nature of the incorporated Sm.\\

The verification of possible Sm$^{2+}$ 4$f$ states in the valence band region is rather difficult due to the intense Eu$^{2+}$ peak at 2 eV binding energy. For a clear detection of Sm$^{2+}$, therefore, the Eu 3$d$ and Sm 3$d$ core levels were measured. Figure \ref{fig:3d} exhibits the core level spectra with the integral background \cite{huefner} subtracted.
The integrated intensity of the peaks has been used to determine the Sm content of Sm$_{\mathrm{x}}$Eu$_{\mathrm{1-x}}$O films, see table \ref{tab:values}. The resulting Eu:Sm ratios fit in general reasonably well with the calculated ratios deduced from the metal fluxes measured using a quartz crystal microbalance and assuming identical re-evaporation rates of Eu and Sm.
We note that for the film with the lowest Sm concentration of 2.4 \% the Sm 3$d$ peak intensity is only of the order of the background. Therefore, this doping level was estimated solely based on the flux ratios.
The shape of the Sm 3$d$ core levels in Sm$_{\mathrm{x}}$Eu$_{\mathrm{1-x}}$O matches well with the ones of the Sm$_2$O$_3$ reference confirming the predominantly trivalent state of Sm in our films. We do not observe any indication for the presence of metallic Sm or Sm$^{2+}$ within the limits of the measurement resolution.\\

\begin{figure}
	\centering
		\includegraphics[width=1\columnwidth]{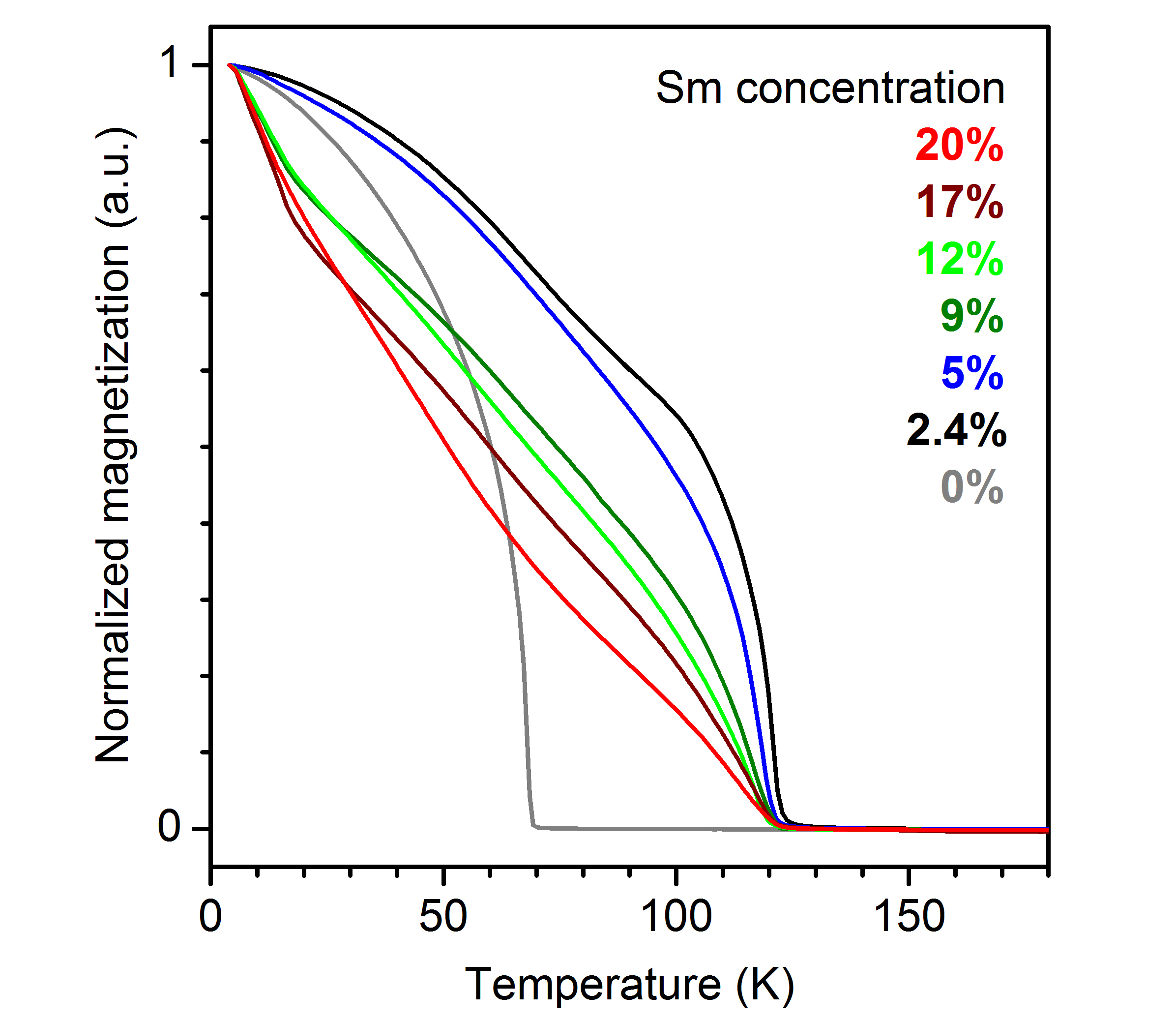}
	\caption{Normalized temperature-dependent magnetization curves of the Sm$_{\mathrm{x}}$Eu$_{\mathrm{1-x}}$O films.}
	\label{fig:MofT}
\end{figure}

In the next step, we investigate the influence of the Sm on the magnetic properties of EuO to verify that the Sm$^{3+}$ ions are successfully integrated into the EuO matrix as indicated by the structural analysis.
The incorporation of trivalent dopants into the EuO lattice is known to enhance the relatively low Curie temperature ($T_C$) of 69 K for the  stoichiometric compound \cite{boyd66}. Examples include doping with trivalent rare earth ions like Gd \cite{ahn68, samokhvalov72, ott06, sutarto09b, melville12}, La \cite{ahn68, schmehl07, melville12, miyazaki10}, Lu \cite{melville12}, and Nd \cite{ahn68} or trivalent transition metals ions like Y \cite{ahn68} and  Sc \cite{altendorf14}.
Indeed, also our  Sm$_{\mathrm{x}}$Eu$_{\mathrm{1-x}}$O  films show enhanced Curie temperatures. The temperature dependent magnetization curves are presented in fig. \ref{fig:MofT}.
The EuO reference sample shows the typical Brillouin like shape with a $T_C$ of 69 K. Upon incorporation of Sm, $T_C$ increases significantly with a maximum of 122 K. The shape of the magnetization curves deviate from a Brillouin function (except perhaps for the 5\% film), indicating the presence of phase separation, i.e., part of the film orders magnetically at 122 K and another part at 69 K or lower.
The observed elevated Curie temperature of the Sm$_{\mathrm{x}}$Eu$_{\mathrm{1-x}}$O films proves the effective integration of samarium as dopant into the EuO lattice and also provides further evidence for the predominantly trivalent character of the incorporated Sm.
In the 1970s, a few studies about the substitution of Eu$^{2+}$ by Sm in EuO polycrystalline samples (Sm:EuO) were published in which a trivalent character of the built-in Sm ions was reported \cite{finkelstein77}, and a higher maximum Curie temperature of about 130 K was found \cite{samokhvalov76, samokhvalov82}.
The magnetization curves of the Sm$_{\mathrm{x}}$Eu$_{\mathrm{1-x}}$O films exhibit a striking similarity with Gd-doped EuO which has a slightly higher maximum T$_C$  of 125 K \cite{sutarto09b}. The 3 K lower T$_C$  value of the Sm-doped films may be explained by the slightly larger ionic radius of Sm$^{3+}$ compared to Gd$^{3+}$, and thus a smaller compression of the lattice, consistent with our previous band structure calculations. \cite{altendorf12}\\

Having unambiguously established that samarium is trivalent in Sm$_{\mathrm{x}}$Eu$_{\mathrm{1-x}}$O, we now address the underlying microscopic reasoning for the observed behavior. Given that the lattice constant for EuO (5.1435 \AA) \cite{McMasters74} is larger than intermediate valence bulk SmO (4.9414 \AA) \cite{krill80}, one would expect that there is enough room to stabilize the larger Sm$^{2+}$ ions in Sm$_{\mathrm{x}}$Eu$_{\mathrm{1-x}}$O.
The observed unexpected 3+ valency can be reasoned as follows. Comparing the lattice constants of $fcc$-NiO (4.17 \AA) \cite{bhatt69} and $fcc$-EuO (5.1435 \AA) \cite{McMasters74}, one can infer that the larger lattice parameter of EuO originates from the fact that the ionic radius of Eu$^{2+}$ is larger than both the  Ni$^{2+}$ and the O$^{2-}$ and that it is the Eu$^{2+}$ ions which determine the lattice constant.
Thus, the O$^{2-}$ ions have enough space to move or 'rattle' in this lattice. This should also be the case for Sm$_{\mathrm{x}}$Eu$_{\mathrm{1-x}}$O, giving the possibility to gain energy by moving the O$^{2-}$ ions closer to the Sm ions so as to form Sm$^{3+}$ having the shorter Sm-O bonds.
A similar observation was also made in La$_{1.5}$Sr$_{0.5}$CoO$_4$, \cite{chang09} in which both Co$^{2+}$ and Co$^{3+}$ ions occupy similar lattice sites.
It was found that the Co$^{3+}$ ions are in the low spin state although the lattice space would be sufficient to stabilize the larger ionic radius - high-spin Co$^{3+}$ ions. \cite{chang09}

\section{Conclusions}
We have prepared high quality single crystalline Sm$_{\mathrm{x}}$Eu$_{\mathrm{1-x}}$O thin films with Sm concentrations up to 20\% \ by  Eu distillation assisted MBE to study the valence state of Sm. Despite the large lattice spacing in the EuO rock salt crystal which is expected to support the divalent state of Sm, XPS and magnetization measurements reveal that the incorporated Sm ions exhibit a predominantly trivalent character. The Curie temperature of the Sm-doped EuO films can be enhanced up to 122 K confirming an effective transfer of charge from the Sm to the unoccupied states of the Eu ions. We infer that the large EuO lattice allows the oxygen ions to move locally in the structure, forming shorter Sm-O bonds and thereby stabilizing the Sm$^{3+}$ valence. 

\begin{acknowledgments}
We thank Ch. Becker and T. Mende for their skillful technical assistance during experiments. The purification of the samarium by P. H\"ohn is gratefully acknowledged.
\end{acknowledgments}


\begin{thebibliography}{38}%
\makeatletter
\providecommand \@ifxundefined [1]{%
 \@ifx{#1\undefined}
}%
\providecommand \@ifnum [1]{%
 \ifnum #1\expandafter \@firstoftwo
 \else \expandafter \@secondoftwo
 \fi
}%
\providecommand \@ifx [1]{%
 \ifx #1\expandafter \@firstoftwo
 \else \expandafter \@secondoftwo
 \fi
}%
\providecommand \natexlab [1]{#1}%
\providecommand \enquote  [1]{``#1''}%
\providecommand \bibnamefont  [1]{#1}%
\providecommand \bibfnamefont [1]{#1}%
\providecommand \citenamefont [1]{#1}%
\providecommand \href@noop [0]{\@secondoftwo}%
\providecommand \href [0]{\begingroup \@sanitize@url \@href}%
\providecommand \@href[1]{\@@startlink{#1}\@@href}%
\providecommand \@@href[1]{\endgroup#1\@@endlink}%
\providecommand \@sanitize@url [0]{\catcode `\\12\catcode `\$12\catcode
  `\&12\catcode `\#12\catcode `\^12\catcode `\_12\catcode `\%12\relax}%
\providecommand \@@startlink[1]{}%
\providecommand \@@endlink[0]{}%
\providecommand \url  [0]{\begingroup\@sanitize@url \@url }%
\providecommand \@url [1]{\endgroup\@href {#1}{\urlprefix }}%
\providecommand \urlprefix  [0]{URL }%
\providecommand \Eprint [0]{\href }%
\providecommand \doibase [0]{http://dx.doi.org/}%
\providecommand \selectlanguage [0]{\@gobble}%
\providecommand \bibinfo  [0]{\@secondoftwo}%
\providecommand \bibfield  [0]{\@secondoftwo}%
\providecommand \translation [1]{[#1]}%
\providecommand \BibitemOpen [0]{}%
\providecommand \bibitemStop [0]{}%
\providecommand \bibitemNoStop [0]{.\EOS\space}%
\providecommand \EOS [0]{\spacefactor3000\relax}%
\providecommand \BibitemShut  [1]{\csname bibitem#1\endcsname}%
\let\auto@bib@innerbib\@empty
%</preamble>
\bibitem [{\citenamefont {Dzero}\ \emph {et~al.}(2010)\citenamefont {Dzero},
  \citenamefont {Sun}, \citenamefont {Galitski},\ and\ \citenamefont
  {Coleman}}]{dzero10}%
  \BibitemOpen
  \bibfield  {author} {\bibinfo {author} {\bibfnamefont {M.}~\bibnamefont
  {Dzero}}, \bibinfo {author} {\bibfnamefont {K.}~\bibnamefont {Sun}}, \bibinfo
  {author} {\bibfnamefont {V.}~\bibnamefont {Galitski}}, \ and\ \bibinfo
  {author} {\bibfnamefont {P.}~\bibnamefont {Coleman}},\ }\href@noop {}
  {\bibfield  {journal} {\bibinfo  {journal} {Phys. Rev. Lett.}\ }\textbf
  {\bibinfo {volume} {104}},\ \bibinfo {pages} {106408} (\bibinfo {year}
  {2010})}\BibitemShut {NoStop}%
\bibitem [{\citenamefont {Takimoto}(2011)}]{takimoto11}%
  \BibitemOpen
  \bibfield  {author} {\bibinfo {author} {\bibfnamefont {T.}~\bibnamefont
  {Takimoto}},\ }\href@noop {} {\bibfield  {journal} {\bibinfo  {journal} {J.
  Phys. Soc. Jpn.}\ }\textbf {\bibinfo {volume} {80}},\ \bibinfo {pages}
  {123710} (\bibinfo {year} {2011})}\BibitemShut {NoStop}%
\bibitem [{\citenamefont {Li}\ \emph {et~al.}(2014)\citenamefont {Li},
  \citenamefont {Li}, \citenamefont {Blaha},\ and\ \citenamefont
  {Kioussis}}]{li14}%
  \BibitemOpen
  \bibfield  {author} {\bibinfo {author} {\bibfnamefont {Z.}~\bibnamefont
  {Li}}, \bibinfo {author} {\bibfnamefont {J.}~\bibnamefont {Li}}, \bibinfo
  {author} {\bibfnamefont {P.}~\bibnamefont {Blaha}}, \ and\ \bibinfo {author}
  {\bibfnamefont {N.}~\bibnamefont {Kioussis}},\ }\href@noop {} {\bibfield
  {journal} {\bibinfo  {journal} {Phys. Rev. B}\ }\textbf {\bibinfo {volume}
  {89}},\ \bibinfo {pages} {121117 (R)} (\bibinfo {year} {2014})}\BibitemShut
  {NoStop}%
\bibitem [{\citenamefont {Kasinathan}\ \emph {et~al.}(2015)\citenamefont
  {Kasinathan}, \citenamefont {Koepernik}, \citenamefont {Tjeng},\ and\
  \citenamefont {Haverkort}}]{kasinathan15}%
  \BibitemOpen
  \bibfield  {author} {\bibinfo {author} {\bibfnamefont {D.}~\bibnamefont
  {Kasinathan}}, \bibinfo {author} {\bibfnamefont {K.}~\bibnamefont
  {Koepernik}}, \bibinfo {author} {\bibfnamefont {L.~H.}\ \bibnamefont
  {Tjeng}}, \ and\ \bibinfo {author} {\bibfnamefont {M.~W.}\ \bibnamefont
  {Haverkort}},\ }\href@noop {} {\bibfield  {journal} {\bibinfo  {journal}
  {Phys. Rev. B}\ }\textbf {\bibinfo {volume} {91}},\ \bibinfo {pages} {195127}
  (\bibinfo {year} {2015})}\BibitemShut {NoStop}%
\bibitem [{\citenamefont {Vainshtein}\ \emph {et~al.}(1965)\citenamefont
  {Vainshtein}, \citenamefont {Blokhin},\ and\ \citenamefont
  {Paderno}}]{vainshtein1965}%
  \BibitemOpen
  \bibfield  {author} {\bibinfo {author} {\bibfnamefont {E.~E.}\ \bibnamefont
  {Vainshtein}}, \bibinfo {author} {\bibfnamefont {S.~M.}\ \bibnamefont
  {Blokhin}}, \ and\ \bibinfo {author} {\bibfnamefont {Y.~B.}\ \bibnamefont
  {Paderno}},\ }\href@noop {} {\bibfield  {journal} {\bibinfo  {journal} {Sov.
  Phys. Solid State}\ }\textbf {\bibinfo {volume} {6}},\ \bibinfo {pages}
  {2318} (\bibinfo {year} {1965})}\BibitemShut {NoStop}%
\bibitem [{\citenamefont {Menth}\ \emph {et~al.}(1969)\citenamefont {Menth},
  \citenamefont {Buehler},\ and\ \citenamefont {Geballe}}]{menth1969}%
  \BibitemOpen
  \bibfield  {author} {\bibinfo {author} {\bibfnamefont {A.}~\bibnamefont
  {Menth}}, \bibinfo {author} {\bibfnamefont {E.}~\bibnamefont {Buehler}}, \
  and\ \bibinfo {author} {\bibfnamefont {T.~H.}\ \bibnamefont {Geballe}},\
  }\href@noop {} {\bibfield  {journal} {\bibinfo  {journal} {Phys. Rev. Lett.}\
  }\textbf {\bibinfo {volume} {22}},\ \bibinfo {pages} {295} (\bibinfo {year}
  {1969})}\BibitemShut {NoStop}%
\bibitem [{\citenamefont {Sidorov}\ \emph {et~al.}(1989)\citenamefont
  {Sidorov}, \citenamefont {Stepanov}, \citenamefont {Khvostantsev},
  \citenamefont {Tsiok}, \citenamefont {Golubkov}, \citenamefont {Oskotski},\
  and\ \citenamefont {Smirnov}}]{sidorov1989}%
  \BibitemOpen
  \bibfield  {author} {\bibinfo {author} {\bibfnamefont {V.~A.}\ \bibnamefont
  {Sidorov}}, \bibinfo {author} {\bibfnamefont {N.~N.}\ \bibnamefont
  {Stepanov}}, \bibinfo {author} {\bibfnamefont {L.~G.}\ \bibnamefont
  {Khvostantsev}}, \bibinfo {author} {\bibfnamefont {O.~B.}\ \bibnamefont
  {Tsiok}}, \bibinfo {author} {\bibfnamefont {A.~V.}\ \bibnamefont {Golubkov}},
  \bibinfo {author} {\bibfnamefont {V.~S.}\ \bibnamefont {Oskotski}}, \ and\
  \bibinfo {author} {\bibfnamefont {I.~A.}\ \bibnamefont {Smirnov}},\
  }\href@noop {} {\bibfield  {journal} {\bibinfo  {journal} {Semicond. Sci.
  Technol.}\ }\textbf {\bibinfo {volume} {4}},\ \bibinfo {pages} {286}
  (\bibinfo {year} {1989})}\BibitemShut {NoStop}%
\bibitem [{\citenamefont {Yacoubi}\ \emph {et~al.}(1979)\citenamefont
  {Yacoubi}, \citenamefont {L{\'e}ger},\ and\ \citenamefont
  {Loriers}}]{yacoubi1979}%
  \BibitemOpen
  \bibfield  {author} {\bibinfo {author} {\bibfnamefont {N.}~\bibnamefont
  {Yacoubi}}, \bibinfo {author} {\bibfnamefont {J.~M.}\ \bibnamefont
  {L{\'e}ger}}, \ and\ \bibinfo {author} {\bibfnamefont {C.}~\bibnamefont
  {Loriers}},\ }\href@noop {} {\bibfield  {journal} {\bibinfo  {journal} {J. de
  Physique Colloques}\ }\textbf {\bibinfo {volume} {40}},\ \bibinfo {pages}
  {C5} (\bibinfo {year} {1979})}\BibitemShut {NoStop}%
\bibitem [{\citenamefont {Peratheepan}\ and\ \citenamefont
  {Strydom}(2015)}]{peratheepan2015}%
  \BibitemOpen
  \bibfield  {author} {\bibinfo {author} {\bibfnamefont {P.}~\bibnamefont
  {Peratheepan}}\ and\ \bibinfo {author} {\bibfnamefont {A.~M.}\ \bibnamefont
  {Strydom}},\ }\href@noop {} {\bibfield  {journal} {\bibinfo  {journal} {J.
  Phys.: Condens. Matter}\ }\textbf {\bibinfo {volume} {27}},\ \bibinfo {pages}
  {095604} (\bibinfo {year} {2015})}\BibitemShut {NoStop}%
\bibitem [{\citenamefont {Arvanitidis}\ \emph {et~al.}(2003)\citenamefont
  {Arvanitidis}, \citenamefont {Papagelis}, \citenamefont {Margadonna},
  \citenamefont {Prassides},\ and\ \citenamefont {Fitch}}]{arvanitidis2003}%
  \BibitemOpen
  \bibfield  {author} {\bibinfo {author} {\bibfnamefont {J.}~\bibnamefont
  {Arvanitidis}}, \bibinfo {author} {\bibfnamefont {K.}~\bibnamefont
  {Papagelis}}, \bibinfo {author} {\bibfnamefont {S.}~\bibnamefont
  {Margadonna}}, \bibinfo {author} {\bibfnamefont {K.}~\bibnamefont
  {Prassides}}, \ and\ \bibinfo {author} {\bibfnamefont {A.~N.}\ \bibnamefont
  {Fitch}},\ }\href@noop {} {\bibfield  {journal} {\bibinfo  {journal}
  {Nature}\ }\textbf {\bibinfo {volume} {425}},\ \bibinfo {pages} {599}
  (\bibinfo {year} {2003})}\BibitemShut {NoStop}%
\bibitem [{\citenamefont {Wertheim}\ and\ \citenamefont
  {Campagna}(1977)}]{wertheim1977}%
  \BibitemOpen
  \bibfield  {author} {\bibinfo {author} {\bibfnamefont {G.~K.}\ \bibnamefont
  {Wertheim}}\ and\ \bibinfo {author} {\bibfnamefont {M.}~\bibnamefont
  {Campagna}},\ }\href@noop {} {\bibfield  {journal} {\bibinfo  {journal}
  {Chem. Phys. Lett.}\ }\textbf {\bibinfo {volume} {47}},\ \bibinfo {pages}
  {182} (\bibinfo {year} {1977})}\BibitemShut {NoStop}%
\bibitem [{\citenamefont {Sutarto}\ \emph
  {et~al.}(2009{\natexlab{a}})\citenamefont {Sutarto}, \citenamefont
  {Altendorf}, \citenamefont {Coloru}, \citenamefont {Moretti~Sala},
  \citenamefont {Haupricht}, \citenamefont {Chang}, \citenamefont {Hu},
  \citenamefont {Sch\"u\ss{}ler-Langeheine}, \citenamefont {Hollmann},
  \citenamefont {Kierspel}, \citenamefont {Hsieh}, \citenamefont {Lin},
  \citenamefont {Chen},\ and\ \citenamefont {Tjeng}}]{sutarto09a}%
  \BibitemOpen
  \bibfield  {author} {\bibinfo {author} {\bibfnamefont {R.}~\bibnamefont
  {Sutarto}}, \bibinfo {author} {\bibfnamefont {S.~G.}\ \bibnamefont
  {Altendorf}}, \bibinfo {author} {\bibfnamefont {B.}~\bibnamefont {Coloru}},
  \bibinfo {author} {\bibfnamefont {M.}~\bibnamefont {Moretti~Sala}}, \bibinfo
  {author} {\bibfnamefont {T.}~\bibnamefont {Haupricht}}, \bibinfo {author}
  {\bibfnamefont {C.~F.}\ \bibnamefont {Chang}}, \bibinfo {author}
  {\bibfnamefont {Z.}~\bibnamefont {Hu}}, \bibinfo {author} {\bibfnamefont
  {C.}~\bibnamefont {Sch\"u\ss{}ler-Langeheine}}, \bibinfo {author}
  {\bibfnamefont {N.}~\bibnamefont {Hollmann}}, \bibinfo {author}
  {\bibfnamefont {H.}~\bibnamefont {Kierspel}}, \bibinfo {author}
  {\bibfnamefont {H.~H.}\ \bibnamefont {Hsieh}}, \bibinfo {author}
  {\bibfnamefont {H.-J.}\ \bibnamefont {Lin}}, \bibinfo {author} {\bibfnamefont
  {C.~T.}\ \bibnamefont {Chen}}, \ and\ \bibinfo {author} {\bibfnamefont
  {L.~H.}\ \bibnamefont {Tjeng}},\ }\href@noop {} {\bibfield  {journal}
  {\bibinfo  {journal} {Phys. Rev. B}\ }\textbf {\bibinfo {volume} {79}},\
  \bibinfo {pages} {205318} (\bibinfo {year} {2009}{\natexlab{a}})}\BibitemShut
  {NoStop}%
\bibitem [{\citenamefont {Leger}\ \emph {et~al.}(1980)\citenamefont {Leger},
  \citenamefont {Yacoubi},\ and\ \citenamefont {Loriers}}]{leger80}%
  \BibitemOpen
  \bibfield  {author} {\bibinfo {author} {\bibfnamefont {J.~M.}\ \bibnamefont
  {Leger}}, \bibinfo {author} {\bibfnamefont {N.}~\bibnamefont {Yacoubi}}, \
  and\ \bibinfo {author} {\bibfnamefont {J.}~\bibnamefont {Loriers}},\
  }\href@noop {} {\bibfield  {journal} {\bibinfo  {journal} {Inorg. Chem.}\
  }\textbf {\bibinfo {volume} {19}},\ \bibinfo {pages} {2252} (\bibinfo {year}
  {1980})}\BibitemShut {NoStop}%
\bibitem [{\citenamefont {Krill}\ \emph {et~al.}(1980)\citenamefont {Krill},
  \citenamefont {Ravet}, \citenamefont {Kappler}, \citenamefont {Abadli},
  \citenamefont {Leger}, \citenamefont {Yacoubi},\ and\ \citenamefont
  {Loriers}}]{krill80}%
  \BibitemOpen
  \bibfield  {author} {\bibinfo {author} {\bibfnamefont {G.}~\bibnamefont
  {Krill}}, \bibinfo {author} {\bibfnamefont {M.~F.}\ \bibnamefont {Ravet}},
  \bibinfo {author} {\bibfnamefont {J.~P.}\ \bibnamefont {Kappler}}, \bibinfo
  {author} {\bibfnamefont {L.}~\bibnamefont {Abadli}}, \bibinfo {author}
  {\bibfnamefont {J.~M.}\ \bibnamefont {Leger}}, \bibinfo {author}
  {\bibfnamefont {N.}~\bibnamefont {Yacoubi}}, \ and\ \bibinfo {author}
  {\bibfnamefont {C.}~\bibnamefont {Loriers}},\ }\href@noop {} {\bibfield
  {journal} {\bibinfo  {journal} {Solid State Commun.}\ }\textbf {\bibinfo
  {volume} {33}},\ \bibinfo {pages} {351} (\bibinfo {year} {1980})}\BibitemShut
  {NoStop}%
\bibitem [{\citenamefont {McMasters}\ \emph {et~al.}(1974)\citenamefont
  {McMasters}, \citenamefont {Gschneidner}, \citenamefont {Kaldis},\ and\
  \citenamefont {Sampietro}}]{McMasters74}%
  \BibitemOpen
  \bibfield  {author} {\bibinfo {author} {\bibfnamefont {O.~D.}\ \bibnamefont
  {McMasters}}, \bibinfo {author} {\bibfnamefont {K.~A.}\ \bibnamefont
  {Gschneidner}}, \bibinfo {author} {\bibfnamefont {E.}~\bibnamefont {Kaldis}},
  \ and\ \bibinfo {author} {\bibfnamefont {G.}~\bibnamefont {Sampietro}},\
  }\href@noop {} {\bibfield  {journal} {\bibinfo  {journal} {J. Chem.
  Thermodyn.}\ }\textbf {\bibinfo {volume} {6}},\ \bibinfo {pages} {845}
  (\bibinfo {year} {1974})}\BibitemShut {NoStop}%
\bibitem [{\citenamefont {Steeneken}\ \emph {et~al.}(2002)\citenamefont
  {Steeneken}, \citenamefont {Tjeng}, \citenamefont {Elfimov}, \citenamefont
  {Sawatzky}, \citenamefont {Ghiringhelli}, \citenamefont {Brookes},\ and\
  \citenamefont {Huang}}]{steeneken02}%
  \BibitemOpen
  \bibfield  {author} {\bibinfo {author} {\bibfnamefont {P.~G.}\ \bibnamefont
  {Steeneken}}, \bibinfo {author} {\bibfnamefont {L.~H.}\ \bibnamefont
  {Tjeng}}, \bibinfo {author} {\bibfnamefont {I.}~\bibnamefont {Elfimov}},
  \bibinfo {author} {\bibfnamefont {G.~A.}\ \bibnamefont {Sawatzky}}, \bibinfo
  {author} {\bibfnamefont {G.}~\bibnamefont {Ghiringhelli}}, \bibinfo {author}
  {\bibfnamefont {N.~B.}\ \bibnamefont {Brookes}}, \ and\ \bibinfo {author}
  {\bibfnamefont {D.-J.}\ \bibnamefont {Huang}},\ }\href@noop {} {\bibfield
  {journal} {\bibinfo  {journal} {Phys. Rev. Lett.}\ }\textbf {\bibinfo
  {volume} {88}},\ \bibinfo {pages} {047201} (\bibinfo {year}
  {2002})}\BibitemShut {NoStop}%
\bibitem [{\citenamefont {Schmehl}\ \emph {et~al.}(2007)\citenamefont
  {Schmehl}, \citenamefont {Vaithyanathan}, \citenamefont {Herrnberger},
  \citenamefont {Thiel}, \citenamefont {Richter}, \citenamefont {Liberati},
  \citenamefont {Heeg}, \citenamefont {R{\"o}ckerath}, \citenamefont
  {Kourkoutis}, \citenamefont {M{\"u}hlbauer}, \citenamefont {B{\"o}ni},
  \citenamefont {Muller}, \citenamefont {Barash}, \citenamefont {Schubert},
  \citenamefont {Idzerda}, \citenamefont {Mannhart},\ and\ \citenamefont
  {Schlom}}]{schmehl07}%
  \BibitemOpen
  \bibfield  {author} {\bibinfo {author} {\bibfnamefont {A.}~\bibnamefont
  {Schmehl}}, \bibinfo {author} {\bibfnamefont {V.}~\bibnamefont
  {Vaithyanathan}}, \bibinfo {author} {\bibfnamefont {A.}~\bibnamefont
  {Herrnberger}}, \bibinfo {author} {\bibfnamefont {S.}~\bibnamefont {Thiel}},
  \bibinfo {author} {\bibfnamefont {C.}~\bibnamefont {Richter}}, \bibinfo
  {author} {\bibfnamefont {M.}~\bibnamefont {Liberati}}, \bibinfo {author}
  {\bibfnamefont {T.}~\bibnamefont {Heeg}}, \bibinfo {author} {\bibfnamefont
  {M.}~\bibnamefont {R{\"o}ckerath}}, \bibinfo {author} {\bibfnamefont {L.~F.}\
  \bibnamefont {Kourkoutis}}, \bibinfo {author} {\bibfnamefont
  {S.}~\bibnamefont {M{\"u}hlbauer}}, \bibinfo {author} {\bibfnamefont
  {P.}~\bibnamefont {B{\"o}ni}}, \bibinfo {author} {\bibfnamefont {D.~A.}\
  \bibnamefont {Muller}}, \bibinfo {author} {\bibfnamefont {Y.}~\bibnamefont
  {Barash}}, \bibinfo {author} {\bibfnamefont {J.}~\bibnamefont {Schubert}},
  \bibinfo {author} {\bibfnamefont {Y.}~\bibnamefont {Idzerda}}, \bibinfo
  {author} {\bibfnamefont {J.}~\bibnamefont {Mannhart}}, \ and\ \bibinfo
  {author} {\bibfnamefont {D.~G.}\ \bibnamefont {Schlom}},\ }\href@noop {}
  {\bibfield  {journal} {\bibinfo  {journal} {Nat. Mater.}\ }\textbf {\bibinfo
  {volume} {6}},\ \bibinfo {pages} {882} (\bibinfo {year} {2007})}\BibitemShut
  {NoStop}%
\bibitem [{\citenamefont {Torrance}\ \emph {et~al.}(1972)\citenamefont
  {Torrance}, \citenamefont {Shafer},\ and\ \citenamefont
  {McGuire}}]{torrance72}%
  \BibitemOpen
  \bibfield  {author} {\bibinfo {author} {\bibfnamefont {J.~B.}\ \bibnamefont
  {Torrance}}, \bibinfo {author} {\bibfnamefont {M.~W.}\ \bibnamefont
  {Shafer}}, \ and\ \bibinfo {author} {\bibfnamefont {T.~R.}\ \bibnamefont
  {McGuire}},\ }\href@noop {} {\bibfield  {journal} {\bibinfo  {journal} {Phys.
  Rev. Lett.}\ }\textbf {\bibinfo {volume} {29}},\ \bibinfo {pages} {1168}
  (\bibinfo {year} {1972})}\BibitemShut {NoStop}%
\bibitem [{\citenamefont {Shapira}\ \emph {et~al.}(1973)\citenamefont
  {Shapira}, \citenamefont {Foner},\ and\ \citenamefont {Reed}}]{shapira73}%
  \BibitemOpen
  \bibfield  {author} {\bibinfo {author} {\bibfnamefont {Y.}~\bibnamefont
  {Shapira}}, \bibinfo {author} {\bibfnamefont {S.}~\bibnamefont {Foner}}, \
  and\ \bibinfo {author} {\bibfnamefont {T.~B.}\ \bibnamefont {Reed}},\
  }\href@noop {} {\bibfield  {journal} {\bibinfo  {journal} {Phys. Rev. B}\
  }\textbf {\bibinfo {volume} {8}},\ \bibinfo {pages} {2299} (\bibinfo {year}
  {1973})}\BibitemShut {NoStop}%
\bibitem [{\citenamefont {Ahn}\ and\ \citenamefont {Shafer}(1970)}]{ahn70b}%
  \BibitemOpen
  \bibfield  {author} {\bibinfo {author} {\bibfnamefont {K.~Y.}\ \bibnamefont
  {Ahn}}\ and\ \bibinfo {author} {\bibfnamefont {M.~W.}\ \bibnamefont
  {Shafer}},\ }\href@noop {} {\bibfield  {journal} {\bibinfo  {journal} {J.
  Appl. Phys.}\ }\textbf {\bibinfo {volume} {41}},\ \bibinfo {pages} {1260}
  (\bibinfo {year} {1970})}\BibitemShut {NoStop}%
\bibitem [{\citenamefont {Wang}\ \emph {et~al.}(1986)\citenamefont {Wang},
  \citenamefont {Schoenes},\ and\ \citenamefont {Kaldis}}]{wang86}%
  \BibitemOpen
  \bibfield  {author} {\bibinfo {author} {\bibfnamefont {H.~Y.}\ \bibnamefont
  {Wang}}, \bibinfo {author} {\bibfnamefont {J.}~\bibnamefont {Schoenes}}, \
  and\ \bibinfo {author} {\bibfnamefont {E.}~\bibnamefont {Kaldis}},\
  }\href@noop {} {\bibfield  {journal} {\bibinfo  {journal} {Helv. Phys. Acta}\
  }\textbf {\bibinfo {volume} {59}},\ \bibinfo {pages} {102} (\bibinfo {year}
  {1986})}\BibitemShut {NoStop}%
\bibitem [{\citenamefont {Strisland}\ \emph {et~al.}(1997)\citenamefont
  {Strisland}, \citenamefont {Raaen}, \citenamefont {Ramstad},\ and\
  \citenamefont {Berg}}]{strisland96}%
  \BibitemOpen
  \bibfield  {author} {\bibinfo {author} {\bibfnamefont {F.}~\bibnamefont
  {Strisland}}, \bibinfo {author} {\bibfnamefont {S.}~\bibnamefont {Raaen}},
  \bibinfo {author} {\bibfnamefont {A.}~\bibnamefont {Ramstad}}, \ and\
  \bibinfo {author} {\bibfnamefont {C.}~\bibnamefont {Berg}},\ }\href@noop {}
  {\bibfield  {journal} {\bibinfo  {journal} {Phys. Rev. B}\ }\textbf {\bibinfo
  {volume} {55}},\ \bibinfo {pages} {1391} (\bibinfo {year}
  {1997})}\BibitemShut {NoStop}%
\bibitem [{\citenamefont {Samokhvalov}\ \emph {et~al.}(1976)\citenamefont
  {Samokhvalov}, \citenamefont {Arbuzova}, \citenamefont {Babushkin},
  \citenamefont {Gizhevskii}, \citenamefont {Efremova}, \citenamefont
  {Simonova},\ and\ \citenamefont {Chebotaev}}]{samokhvalov76}%
  \BibitemOpen
  \bibfield  {author} {\bibinfo {author} {\bibfnamefont {A.~A.}\ \bibnamefont
  {Samokhvalov}}, \bibinfo {author} {\bibfnamefont {T.~I.}\ \bibnamefont
  {Arbuzova}}, \bibinfo {author} {\bibfnamefont {V.~S.}\ \bibnamefont
  {Babushkin}}, \bibinfo {author} {\bibfnamefont {B.~A.}\ \bibnamefont
  {Gizhevskii}}, \bibinfo {author} {\bibfnamefont {N.~N.}\ \bibnamefont
  {Efremova}}, \bibinfo {author} {\bibfnamefont {M.~I.}\ \bibnamefont
  {Simonova}}, \ and\ \bibinfo {author} {\bibfnamefont {N.~M.}\ \bibnamefont
  {Chebotaev}},\ }\href@noop {} {\bibfield  {journal} {\bibinfo  {journal}
  {Sov. Phys. Solid State}\ }\textbf {\bibinfo {volume} {18}},\ \bibinfo
  {pages} {1655} (\bibinfo {year} {1976})}\BibitemShut {NoStop}%
\bibitem [{\citenamefont {H{\"u}fner}(2003)}]{huefner}%
  \BibitemOpen
  \bibfield  {author} {\bibinfo {author} {\bibfnamefont {S.}~\bibnamefont
  {H{\"u}fner}},\ }in\ \href@noop {} {\emph {\bibinfo {booktitle}
  {Photoelectron Spectroscopy}}}\ (\bibinfo  {publisher} {Springer},\ \bibinfo
  {year} {2003})\BibitemShut {NoStop}%
\bibitem [{\citenamefont {Wertheim}\ and\ \citenamefont
  {Crecelius}(1978)}]{wertheim78}%
  \BibitemOpen
  \bibfield  {author} {\bibinfo {author} {\bibfnamefont {G.~K.}\ \bibnamefont
  {Wertheim}}\ and\ \bibinfo {author} {\bibfnamefont {G.}~\bibnamefont
  {Crecelius}},\ }\href@noop {} {\bibfield  {journal} {\bibinfo  {journal}
  {Phys. Rev. Lett.}\ }\textbf {\bibinfo {volume} {40}},\ \bibinfo {pages}
  {813} (\bibinfo {year} {1978})}\BibitemShut {NoStop}%
\bibitem [{\citenamefont {Boyd}(1966)}]{boyd66}%
  \BibitemOpen
  \bibfield  {author} {\bibinfo {author} {\bibfnamefont {E.~L.}\ \bibnamefont
  {Boyd}},\ }\href@noop {} {\bibfield  {journal} {\bibinfo  {journal} {Phys.
  Rev.}\ }\textbf {\bibinfo {volume} {145}},\ \bibinfo {pages} {174} (\bibinfo
  {year} {1966})}\BibitemShut {NoStop}%
\bibitem [{\citenamefont {Ahn}\ and\ \citenamefont {McGuire}(1968)}]{ahn68}%
  \BibitemOpen
  \bibfield  {author} {\bibinfo {author} {\bibfnamefont {K.~Y.}\ \bibnamefont
  {Ahn}}\ and\ \bibinfo {author} {\bibfnamefont {T.~R.}\ \bibnamefont
  {McGuire}},\ }\href@noop {} {\bibfield  {journal} {\bibinfo  {journal} {J.
  Appl. Phys.}\ }\textbf {\bibinfo {volume} {39}},\ \bibinfo {pages} {5061}
  (\bibinfo {year} {1968})}\BibitemShut {NoStop}%
\bibitem [{\citenamefont {Samokhvalov}\ \emph {et~al.}(1972)\citenamefont
  {Samokhvalov}, \citenamefont {Gizhevskii}, \citenamefont {Simonova},\ and\
  \citenamefont {Solin}}]{samokhvalov72}%
  \BibitemOpen
  \bibfield  {author} {\bibinfo {author} {\bibfnamefont {A.~A.}\ \bibnamefont
  {Samokhvalov}}, \bibinfo {author} {\bibfnamefont {B.~A.}\ \bibnamefont
  {Gizhevskii}}, \bibinfo {author} {\bibfnamefont {M.~I.}\ \bibnamefont
  {Simonova}}, \ and\ \bibinfo {author} {\bibfnamefont {N.~I.}\ \bibnamefont
  {Solin}},\ }\href@noop {} {\bibfield  {journal} {\bibinfo  {journal} {Sov.
  Phys. Solid State}\ }\textbf {\bibinfo {volume} {14}},\ \bibinfo {pages}
  {230} (\bibinfo {year} {1972})}\BibitemShut {NoStop}%
\bibitem [{\citenamefont {Ott}\ \emph {et~al.}(2006)\citenamefont {Ott},
  \citenamefont {Heise}, \citenamefont {Sutarto}, \citenamefont {Hu},
  \citenamefont {Chang}, \citenamefont {Hsieh}, \citenamefont {Lin},
  \citenamefont {Chen},\ and\ \citenamefont {Tjeng}}]{ott06}%
  \BibitemOpen
  \bibfield  {author} {\bibinfo {author} {\bibfnamefont {H.}~\bibnamefont
  {Ott}}, \bibinfo {author} {\bibfnamefont {S.~J.}\ \bibnamefont {Heise}},
  \bibinfo {author} {\bibfnamefont {R.}~\bibnamefont {Sutarto}}, \bibinfo
  {author} {\bibfnamefont {Z.}~\bibnamefont {Hu}}, \bibinfo {author}
  {\bibfnamefont {C.~F.}\ \bibnamefont {Chang}}, \bibinfo {author}
  {\bibfnamefont {H.~H.}\ \bibnamefont {Hsieh}}, \bibinfo {author}
  {\bibfnamefont {H.-J.}\ \bibnamefont {Lin}}, \bibinfo {author} {\bibfnamefont
  {C.~T.}\ \bibnamefont {Chen}}, \ and\ \bibinfo {author} {\bibfnamefont
  {L.~H.}\ \bibnamefont {Tjeng}},\ }\href@noop {} {\bibfield  {journal}
  {\bibinfo  {journal} {Phys. Rev. B}\ }\textbf {\bibinfo {volume} {73}},\
  \bibinfo {pages} {094407} (\bibinfo {year} {2006})}\BibitemShut {NoStop}%
\bibitem [{\citenamefont {Sutarto}\ \emph
  {et~al.}(2009{\natexlab{b}})\citenamefont {Sutarto}, \citenamefont
  {Altendorf}, \citenamefont {Coloru}, \citenamefont {Moretti~Sala},
  \citenamefont {Haupricht}, \citenamefont {Chang}, \citenamefont {Hu},
  \citenamefont {Sch\"u\ss{}ler-Langeheine}, \citenamefont {Hollmann},
  \citenamefont {Kierspel}, \citenamefont {Mydosh}, \citenamefont {Hsieh},
  \citenamefont {Lin}, \citenamefont {Chen},\ and\ \citenamefont
  {Tjeng}}]{sutarto09b}%
  \BibitemOpen
  \bibfield  {author} {\bibinfo {author} {\bibfnamefont {R.}~\bibnamefont
  {Sutarto}}, \bibinfo {author} {\bibfnamefont {S.~G.}\ \bibnamefont
  {Altendorf}}, \bibinfo {author} {\bibfnamefont {B.}~\bibnamefont {Coloru}},
  \bibinfo {author} {\bibfnamefont {M.}~\bibnamefont {Moretti~Sala}}, \bibinfo
  {author} {\bibfnamefont {T.}~\bibnamefont {Haupricht}}, \bibinfo {author}
  {\bibfnamefont {C.~F.}\ \bibnamefont {Chang}}, \bibinfo {author}
  {\bibfnamefont {Z.}~\bibnamefont {Hu}}, \bibinfo {author} {\bibfnamefont
  {C.}~\bibnamefont {Sch\"u\ss{}ler-Langeheine}}, \bibinfo {author}
  {\bibfnamefont {N.}~\bibnamefont {Hollmann}}, \bibinfo {author}
  {\bibfnamefont {H.}~\bibnamefont {Kierspel}}, \bibinfo {author}
  {\bibfnamefont {J.~A.}\ \bibnamefont {Mydosh}}, \bibinfo {author}
  {\bibfnamefont {H.~H.}\ \bibnamefont {Hsieh}}, \bibinfo {author}
  {\bibfnamefont {H.-J.}\ \bibnamefont {Lin}}, \bibinfo {author} {\bibfnamefont
  {C.~T.}\ \bibnamefont {Chen}}, \ and\ \bibinfo {author} {\bibfnamefont
  {L.~H.}\ \bibnamefont {Tjeng}},\ }\href@noop {} {\bibfield  {journal}
  {\bibinfo  {journal} {Phys. Rev. B}\ }\textbf {\bibinfo {volume} {80}},\
  \bibinfo {pages} {085308} (\bibinfo {year} {2009}{\natexlab{b}})}\BibitemShut
  {NoStop}%
\bibitem [{\citenamefont {Melville}\ \emph {et~al.}(2012)\citenamefont
  {Melville}, \citenamefont {Mairoser}, \citenamefont {Schmehl}, \citenamefont
  {Shai}, \citenamefont {Monkman}, \citenamefont {Harter}, \citenamefont
  {Heeg}, \citenamefont {Holl\"ander}, \citenamefont {Schubert}, \citenamefont
  {Shen}, \citenamefont {Mannhart},\ and\ \citenamefont {Schlom}}]{melville12}%
  \BibitemOpen
  \bibfield  {author} {\bibinfo {author} {\bibfnamefont {A.}~\bibnamefont
  {Melville}}, \bibinfo {author} {\bibfnamefont {T.}~\bibnamefont {Mairoser}},
  \bibinfo {author} {\bibfnamefont {A.}~\bibnamefont {Schmehl}}, \bibinfo
  {author} {\bibfnamefont {D.~E.}\ \bibnamefont {Shai}}, \bibinfo {author}
  {\bibfnamefont {E.~J.}\ \bibnamefont {Monkman}}, \bibinfo {author}
  {\bibfnamefont {J.~W.}\ \bibnamefont {Harter}}, \bibinfo {author}
  {\bibfnamefont {T.}~\bibnamefont {Heeg}}, \bibinfo {author} {\bibfnamefont
  {B.}~\bibnamefont {Holl\"ander}}, \bibinfo {author} {\bibfnamefont
  {J.}~\bibnamefont {Schubert}}, \bibinfo {author} {\bibfnamefont {K.~M.}\
  \bibnamefont {Shen}}, \bibinfo {author} {\bibfnamefont {J.}~\bibnamefont
  {Mannhart}}, \ and\ \bibinfo {author} {\bibfnamefont {D.~G.}\ \bibnamefont
  {Schlom}},\ }\href@noop {} {\bibfield  {journal} {\bibinfo  {journal} {Appl.
  Phys. Lett.}\ }\textbf {\bibinfo {volume} {100}},\ \bibinfo {pages} {222101}
  (\bibinfo {year} {2012})}\BibitemShut {NoStop}%
\bibitem [{\citenamefont {Miyazaki}\ \emph {et~al.}(2010)\citenamefont
  {Miyazaki}, \citenamefont {Im}, \citenamefont {Terashima}, \citenamefont
  {Yagi}, \citenamefont {Kato}, \citenamefont {Soda}, \citenamefont {Ito},\
  and\ \citenamefont {Kimura}}]{miyazaki10}%
  \BibitemOpen
  \bibfield  {author} {\bibinfo {author} {\bibfnamefont {H.}~\bibnamefont
  {Miyazaki}}, \bibinfo {author} {\bibfnamefont {H.~J.}\ \bibnamefont {Im}},
  \bibinfo {author} {\bibfnamefont {K.}~\bibnamefont {Terashima}}, \bibinfo
  {author} {\bibfnamefont {S.}~\bibnamefont {Yagi}}, \bibinfo {author}
  {\bibfnamefont {M.}~\bibnamefont {Kato}}, \bibinfo {author} {\bibfnamefont
  {K.}~\bibnamefont {Soda}}, \bibinfo {author} {\bibfnamefont {T.}~\bibnamefont
  {Ito}}, \ and\ \bibinfo {author} {\bibfnamefont {S.}~\bibnamefont {Kimura}},\
  }\href@noop {} {\bibfield  {journal} {\bibinfo  {journal} {Appl. Phys.
  Lett.}\ }\textbf {\bibinfo {volume} {96}},\ \bibinfo {pages} {232503}
  (\bibinfo {year} {2010})}\BibitemShut {NoStop}%
\bibitem [{\citenamefont {Altendorf}\ \emph {et~al.}(2014)\citenamefont
  {Altendorf}, \citenamefont {Reisner}, \citenamefont {Chang}, \citenamefont
  {Hollmann}, \citenamefont {Rata},\ and\ \citenamefont {Tjeng}}]{altendorf14}%
  \BibitemOpen
  \bibfield  {author} {\bibinfo {author} {\bibfnamefont {S.~G.}\ \bibnamefont
  {Altendorf}}, \bibinfo {author} {\bibfnamefont {A.}~\bibnamefont {Reisner}},
  \bibinfo {author} {\bibfnamefont {C.~F.}\ \bibnamefont {Chang}}, \bibinfo
  {author} {\bibfnamefont {N.}~\bibnamefont {Hollmann}}, \bibinfo {author}
  {\bibfnamefont {A.~D.}\ \bibnamefont {Rata}}, \ and\ \bibinfo {author}
  {\bibfnamefont {L.~H.}\ \bibnamefont {Tjeng}},\ }\href@noop {} {\bibfield
  {journal} {\bibinfo  {journal} {Appl. Phys. Lett.}\ }\textbf {\bibinfo
  {volume} {104}},\ \bibinfo {pages} {052403} (\bibinfo {year}
  {2014})}\BibitemShut {NoStop}%
\bibitem [{\citenamefont {Finkelstein}\ \emph {et~al.}(1977)\citenamefont
  {Finkelstein}, \citenamefont {Efremova}, \citenamefont {Simonova},
  \citenamefont {Lobachevskaya}, \citenamefont {Samokhvalov}, \citenamefont
  {Bamburov},\ and\ \citenamefont {Nemnonov}}]{finkelstein77}%
  \BibitemOpen
  \bibfield  {author} {\bibinfo {author} {\bibfnamefont {L.~D.}\ \bibnamefont
  {Finkelstein}}, \bibinfo {author} {\bibfnamefont {N.~N.}\ \bibnamefont
  {Efremova}}, \bibinfo {author} {\bibfnamefont {M.~I.}\ \bibnamefont
  {Simonova}}, \bibinfo {author} {\bibfnamefont {N.~I.}\ \bibnamefont
  {Lobachevskaya}}, \bibinfo {author} {\bibfnamefont {A.~A.}\ \bibnamefont
  {Samokhvalov}}, \bibinfo {author} {\bibfnamefont {V.~G.}\ \bibnamefont
  {Bamburov}}, \ and\ \bibinfo {author} {\bibfnamefont {S.~A.}\ \bibnamefont
  {Nemnonov}},\ }\href@noop {} {\bibfield  {journal} {\bibinfo  {journal} {Sov.
  Phys. Solid State}\ }\textbf {\bibinfo {volume} {19}},\ \bibinfo {pages}
  {2168} (\bibinfo {year} {1977})}\BibitemShut {NoStop}%
\bibitem [{\citenamefont {Samokhvalov}\ \emph {et~al.}(1982)\citenamefont
  {Samokhvalov}, \citenamefont {Gizhevskii}, \citenamefont {Simonova},
  \citenamefont {Chebotaev},\ and\ \citenamefont
  {Falkovskaya}}]{samokhvalov82}%
  \BibitemOpen
  \bibfield  {author} {\bibinfo {author} {\bibfnamefont {A.~A.}\ \bibnamefont
  {Samokhvalov}}, \bibinfo {author} {\bibfnamefont {B.~A.}\ \bibnamefont
  {Gizhevskii}}, \bibinfo {author} {\bibfnamefont {M.~I.}\ \bibnamefont
  {Simonova}}, \bibinfo {author} {\bibfnamefont {N.~M.}\ \bibnamefont
  {Chebotaev}}, \ and\ \bibinfo {author} {\bibfnamefont {L.~D.}\ \bibnamefont
  {Falkovskaya}},\ }\href@noop {} {\bibfield  {journal} {\bibinfo  {journal}
  {Sov. Phys. Solid State}\ }\textbf {\bibinfo {volume} {24}},\ \bibinfo
  {pages} {1103} (\bibinfo {year} {1982})}\BibitemShut {NoStop}%
\bibitem [{\citenamefont {Altendorf}\ \emph {et~al.}(2012)\citenamefont
  {Altendorf}, \citenamefont {Hollmann}, \citenamefont {Sutarto}, \citenamefont
  {Caspers}, \citenamefont {Wicks}, \citenamefont {Chin}, \citenamefont {Hu},
  \citenamefont {Kierspel}, \citenamefont {Elfimov}, \citenamefont {Hsieh},
  \citenamefont {Lin}, \citenamefont {Chen},\ and\ \citenamefont
  {Tjeng}}]{altendorf12}%
  \BibitemOpen
  \bibfield  {author} {\bibinfo {author} {\bibfnamefont {S.~G.}\ \bibnamefont
  {Altendorf}}, \bibinfo {author} {\bibfnamefont {N.}~\bibnamefont {Hollmann}},
  \bibinfo {author} {\bibfnamefont {R.}~\bibnamefont {Sutarto}}, \bibinfo
  {author} {\bibfnamefont {C.}~\bibnamefont {Caspers}}, \bibinfo {author}
  {\bibfnamefont {R.~C.}\ \bibnamefont {Wicks}}, \bibinfo {author}
  {\bibfnamefont {Y.-Y.}\ \bibnamefont {Chin}}, \bibinfo {author}
  {\bibfnamefont {Z.}~\bibnamefont {Hu}}, \bibinfo {author} {\bibfnamefont
  {H.}~\bibnamefont {Kierspel}}, \bibinfo {author} {\bibfnamefont {I.~S.}\
  \bibnamefont {Elfimov}}, \bibinfo {author} {\bibfnamefont {H.~H.}\
  \bibnamefont {Hsieh}}, \bibinfo {author} {\bibfnamefont {H.-J.}\ \bibnamefont
  {Lin}}, \bibinfo {author} {\bibfnamefont {C.~T.}\ \bibnamefont {Chen}}, \
  and\ \bibinfo {author} {\bibfnamefont {L.~H.}\ \bibnamefont {Tjeng}},\
  }\href@noop {} {\bibfield  {journal} {\bibinfo  {journal} {Phys. Rev. B}\
  }\textbf {\bibinfo {volume} {85}},\ \bibinfo {pages} {081201} (\bibinfo
  {year} {2012})}\BibitemShut {NoStop}%
\bibitem [{\citenamefont {Bhatt}\ and\ \citenamefont
  {Merchant}(1969)}]{bhatt69}%
  \BibitemOpen
  \bibfield  {author} {\bibinfo {author} {\bibfnamefont {S.~J.}\ \bibnamefont
  {Bhatt}}\ and\ \bibinfo {author} {\bibfnamefont {H.~D.}\ \bibnamefont
  {Merchant}},\ }\href@noop {} {\bibfield  {journal} {\bibinfo  {journal} {J.
  Am. Ceram. Soc.}\ }\textbf {\bibinfo {volume} {52}},\ \bibinfo {pages} {452}
  (\bibinfo {year} {1969})}\BibitemShut {NoStop}%
\bibitem [{\citenamefont {Chang}\ \emph {et~al.}(2009)\citenamefont {Chang},
  \citenamefont {Hu}, \citenamefont {Wu}, \citenamefont {Burnus}, \citenamefont
  {Hollmann}, \citenamefont {Benomar}, \citenamefont {Lorenz}, \citenamefont
  {Tanaka}, \citenamefont {Lin}, \citenamefont {Hsieh}, \citenamefont {Chen},\
  and\ \citenamefont {Tjeng}}]{chang09}%
  \BibitemOpen
  \bibfield  {author} {\bibinfo {author} {\bibfnamefont {C.~F.}\ \bibnamefont
  {Chang}}, \bibinfo {author} {\bibfnamefont {Z.}~\bibnamefont {Hu}}, \bibinfo
  {author} {\bibfnamefont {H.}~\bibnamefont {Wu}}, \bibinfo {author}
  {\bibfnamefont {T.}~\bibnamefont {Burnus}}, \bibinfo {author} {\bibfnamefont
  {N.}~\bibnamefont {Hollmann}}, \bibinfo {author} {\bibfnamefont
  {M.}~\bibnamefont {Benomar}}, \bibinfo {author} {\bibfnamefont
  {T.}~\bibnamefont {Lorenz}}, \bibinfo {author} {\bibfnamefont
  {A.}~\bibnamefont {Tanaka}}, \bibinfo {author} {\bibfnamefont {H.-J.}\
  \bibnamefont {Lin}}, \bibinfo {author} {\bibfnamefont {H.~H.}\ \bibnamefont
  {Hsieh}}, \bibinfo {author} {\bibfnamefont {C.~T.}\ \bibnamefont {Chen}}, \
  and\ \bibinfo {author} {\bibfnamefont {L.~H.}\ \bibnamefont {Tjeng}},\
  }\href@noop {} {\bibfield  {journal} {\bibinfo  {journal} {Phys. Rev. Lett.}\
  }\textbf {\bibinfo {volume} {102}},\ \bibinfo {pages} {116401} (\bibinfo
  {year} {2009})}\BibitemShut {NoStop}%
\end{thebibliography}
\end{document}